\documentclass[aps,pra,twocolumn,amsmath,amssymb,superscriptaddress,longbibliography]{revtex4-1}
\usepackage[english]{babel}
\usepackage{amssymb}
\usepackage{dcolumn}
\usepackage{bm}
\usepackage{graphicx}
\usepackage{amsmath}
\usepackage{graphicx}        
\usepackage{rotating}
\graphicspath{{pict/}{}}

\usepackage{dcolumn}
\usepackage{bm}
\usepackage[pdfstartview=Fit, CJKbookmarks=true, bookmarksnumbered=true, bookmarksopen=true, colorlinks=true, pdfborder=001, citecolor=blue, linkcolor=blue, urlcolor=blue, linktocpage=true] {hyperref}

\begin{document}

\title{Strongly tilted field induced fractional quantized-drift in non-interacting system}

\author{Bo Zhu}
\affiliation{Key Laboratory of Low-Dimension Quantum Structures and Quantum Control of Ministry of Education, Synergetic Innovation Center for Quantum Effects and Applications, and Department of Physics, Hunan Normal University, Changsha 410081, China}
\affiliation{Institute of Mathematics and Physics, Central South University of Forestry and Technology, Changsha 410004, China}

\author{Zhi Tan}
\affiliation{Institute of Quantum Precision Measurement, State Key Laboratory of Radio Frequency Heterogeneous Integration, College of Physics and Optoelectronic Engineering, Shenzhen University, Shenzhen 518060, China}

\author{Huilin Gong}
\affiliation{Institute of Mathematics and Physics, Central South University of Forestry and Technology, Changsha 410004, China}

\author{Honghua Zhong}
\altaffiliation{hhzhong115@163.com.}
\affiliation{Key Laboratory of Low-Dimension Quantum Structures and Quantum Control of Ministry of Education, Synergetic Innovation Center for Quantum Effects and Applications, and Department of Physics, Hunan Normal University, Changsha 410081, China}

\author{Xin-You L\"{u}}
\affiliation{School of Physics and Institute for Quantum Science and Engineering, Huazhong University of Science and Technology, and Wuhan Institute of Quantum Technology, Wuhan 430074, China}
\author{Xiaoguang Wang}
\affiliation{Key Laboratory of Quantum State and Optical Field Manipulation of Zhejiang Province, Department of Physics, Zhejiang Sci-Tech University, Hangzhou 310018, China}

\date{\today}

\begin{abstract}
Fractional quantized response appears to be a distinctive characteristic in interacting topological systems. Here, we discover a novel phenomenon of tilt-induced fractional
quantize drift in non-interacting system constructed by a time-modulated superlattice subjected to a external time-independent gradient potential.
Depending on the tilt strength, Rabi oscillations between adjacent lowest enegy bands caused
by Landau-Zener tunneling, can induce that the one-cycle-averaged drift displacement is fraction,
which is relate to the ratio of the sum of Chern numbers of multiple bands
to the number of energy bands involved in Landau Zener tunneling. As representative examples, we construct fractional (1/3, 1/2) quantize drift only via adjusting period of lattice. The numerical simulations allow us to consider a realistic setup amenable of an experimental realization.
Our findings will expand the research implications of both fractional quantize response and topological materials.

\end{abstract}

\maketitle

Fractional quantum Hall effect, where the interaction between particles leads to fractionally quantized Hall conductance, has attracted much attention in several physical fields ranging from condensed matter physics to optics~\cite{tsui1982two, laughlin1983anomalous, haldane1983fractional, jain1989composite, grusdt2016interferometric, umucalilar2018time, macaluso2020charge, repellin2019detecting, kalmeyer1987equivalence, rosson2019bosonic, wang2011fractional, wang2011fractional,qi2011generic, benedict1999theory, girvin1986magneto, kol1993fractional, hafezi2007fractional, 2005Fractional, parameswaran2013fractional, bergholtz2013topological, yoshida2020fate, knuppel2019nonlinear, clark2020observation, roushan2017chiral, anderson2016engineering}.
As lattice versions of the fractional quantum Hall effect, the concept of fractional Chern insulators (FCIs) has been introduced in
cold atom systems~\cite{ravciunas2018creating, aidelsburger2013realization, cooper2013reaching, grusdt2014realization, he2017realizing, motruk2017phase, repellin2017creating, hudomal2019bosonic, palmer2006high, kjall2012edge, luo2013edge, taddia2017topological, dong2018charge, ravciunas2018creating}.
In view of realizing strongly correlated topological phases of ultracold atoms in optical lattices~\cite{goldman2016topological, cooper2019topological, de2019observation}.
This progress should soon lead to the realization of FCIs in small interacting atomic systems  ~\cite{weber2022experimentally, zhao2023fractional}.
There, the interaction between atoms played the key role, giving rise to the formation of fractional quantum Hall-type states.
The fractional quantum  Hall response has been measured in weakly interacting gases through various probes, including center-of-mass drifts \cite{2020Fractional, 2020Detecting}.
A conceptually different approach has been taken by treating the interactions of many atoms or photons in the mean-field limit using nonlinearity.
This approach has led to the prediction and observation of fractional Thouless pumping of solitons~\cite{jurgensen2023quantized, fu2022nonlinear, fu2022nonlinear, fu2022two, jurgensen2023quantized, tao2024nonlinearity, tao2025nonlinearity, Bohm2025Quantum}.
Yet, it is still unclear whether the fractional quantum Hall response can be extracted and used as a topological marker in non-interacting systems.

In this Letter, we address tilt-induced fractional
quantize drift for a quantum particle tapped in an optical superlattice created
by two lattices subjected to a external time-independent gradient potential. Our two main findings are as follows: First, there exists
a threshold area of tilt strength, below which the transverse
drift displacement of dynamical evolution in real space for wave packet is integral. Above threshold area, the
drift displacement becomes fractional, acquires direction is related to the reduced Chern number defined by line integral of Berry curvature. Second, tilt-induced fractional
quantize drift can well be described by the energy band theory. They occur due to the tilt-induced Rabi oscillations of particle between lowest bands with different Chern numbers caused by Landau-Zener tunneling.
The sum of Chern numbers of multiple bands determines the direction and fractional magnitude of one-cycle-averaged drift displacement.
In addition, due to the tilt automatically cancels the group velocity from energy dispersion, the optional state with any particular momentum in a band can be chosen as the initial state in wave
packet dynamics. The relation of fractional quantized response of
non-interacting systems with Landau-Zener tunneling of multiple bands is established for the first time.

We start by considering a quantum particle in time-modulated superlattice subjected to a external time-independent gradient potential, which is described by the following Hamiltonian
\begin{eqnarray} \label{Eq1} H=-\frac{1}{2}\partial^2_x+V_0(x,t)+V_1(x).
\end{eqnarray}
Here,
coordinate $x$ and time $t$ are measured in the units $\Lambda/\pi$ and $\hbar/2E_r$, respectively,
in which $E_r=\hbar^2 \pi^2 /(2\Lambda^2m)$ is the recoil energy,  $\Lambda$ is a characteristic length defining the period of the superlattice potential $V_0(x,t)$ and $m$ is the particle mass.
$V_1(x)=Fx$ is gradient potential, and $F$ describes the magnitude of the tilt, which can be achieved by applying a magnetic field gradient or aligning the superlattice with gravity.
The superlattice potential is modeled by
\begin{eqnarray} \label{Eq2}
V_0(x,t)=-\tau_1\cos^2(\pi x/d_1)-\tau_2\cos^2(\pi x /d_2-\omega t),
\end{eqnarray}
where $\tau_{1,2}$ and $d_{1,2}$ are the dimensionless depths and  periods and  of the constitutive lattices.
The first stationary lattice can be created by two counter-propagating monochromatic laser beams~\cite{lignier2007dynamical, zenesini2010c, morsch2006dynamics, fu2022nonlinear}, while the second lattice moving with the dimensionless velocity $\omega_L=\omega d_2/\pi$ is created by two counterpropagating beams with the frequency detuning $\sim \omega$~\cite{morsch2006dynamics, fallani2003optically, fu2022nonlinear}.
We require $\omega\ll1$ to be a small parameter determining adiabatic displacement of the second lattice.
The periods $d_1$ and $d_2$ are commensurate $d_1/d_2=n_1/n_2$, where $n_1$ and $n_2$ are co-prime integers,
and thus it results in an overall period $L=n_1d_2=n_2d_1$ in the superlattice potential $V_0(x,t)$.
\begin{figure}[htp]
\center
\includegraphics[width=\columnwidth]{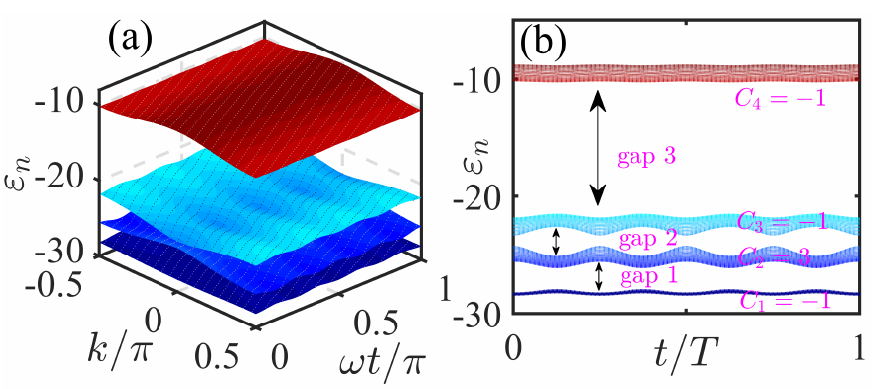}
\caption{(a) The complete two-dimensional energy band in the first Brillouin zone, where thin solid line represents the uniform sampling of the two-dimensional energy bands in the presence of tilt. (b) The equivalent one-dimensional time-dependent energy bands. The other parameters are chosen as $\tau_1=\tau_2=25$, $d_1=1/2, d_2=2/3$, $w=0.01$ and $\xi=3$.
}\label{Fig1s}
\end{figure}

In the second quantization theory,
the instantaneous Hamiltonian in momentum space $\hat{H}(k,t)$ can be given as,
\begin{small}
\begin{eqnarray} \label{Ham5}
\hat{H}(k,t)&=&-\sum_j\left[(\frac{\tau_2}{4}e^{i 2 \omega t}) \hat{c}_j^{\dagger} \hat{c}_{j-\frac{L}{d_2}}+(\frac{\tau_1}{4}) \hat{c}_j^{\dagger} \hat{c}_{j-\frac{L}{d_1}}+\text { H.c. }\right] \nonumber \\
&+&\frac{1}{2}\sum_j\left[\left(k-Ft+\frac{2 \pi}{L} j\right)^2-\left(\tau_1+\tau_2\right)\right] \hat{n}_j.
\end{eqnarray}
\end{small}
where $\hat{c}^{(\dagger)}_j$ is the annihilation (creation) operator of lattice $j$, and $k\in[-\pi/L,\pi/L]$ denotes quasi-momentum.
If we replace ($k-Ft$) and $\omega t$ with $K_x$ and $K_y$ respectively, the Hamiltonian $\hat{H}(k,t)$ can be regarded as an equivalent static Hamiltonian $\hat{H}(K_x,K_y)$,
\begin{small}
\begin{eqnarray} \label{Ham7}
\hat{H}(K_x,K_y)&=&-\sum_j\left[(\frac{\tau_2}{4}e^{i 2 K_y}) \hat{c}_j^{\dagger} \hat{c}_{j-\frac{L}{d_2}}+(\frac{\tau_1}{4}) \hat{c}_j^{\dagger} \hat{c}_{j-\frac{L}{d_1}}
+\text { H.c. }\right] \nonumber \\
&+&\frac{1}{2}\sum_j\left[\left(K_x+\frac{2 \pi}{L} j\right)^2-\left(\tau_1+\tau_2\right)\right] \hat{n}_j,
\end{eqnarray}
\end{small}
where $K_x\in[-\pi/L,\pi/L]$ and $K_y\in[0,\pi]$ can be understood as the quasi-momenta of the two-dimensional first Brillouin zone (See section A of supplemental material).
The conventional Chern number of energy band can be easily calculated based on the static Hamiltonian.
However, in our system, the introduction of tilt results in both $K_x$ and $K_y$ becoming time-dependent parameters, whose period is given by $T_F=2\pi/(LF)$ and $T_m=\pi/\omega$, respectively.
Thus, if we choose any initial momentum $k=k_0$, the instantaneous energy will uniformly sample the entire two-dimensional parameter space $(K_x, K_y)$, and its sampling density is closely related to the tilt strength $F$ and frequency $\omega$.
Especially, when $T_m/T_F\rightarrow0$ or $\gg1$, the topological properties of the two-dimensional energy bands characterized by the two-dimensional parameter space $(K_x, K_y)$ can be replaced by one-dimensional time-dependent energy bands.
For simplicity, we assume $T_m/T_F=p/q$, where $p$ and $q$ are co-prime integers. This results in an overall time period   $T=qT_m=pT_F$,
in which we choose $q=1$ in this article.

\begin{figure*}[!t]
\center
\includegraphics[width=7in]{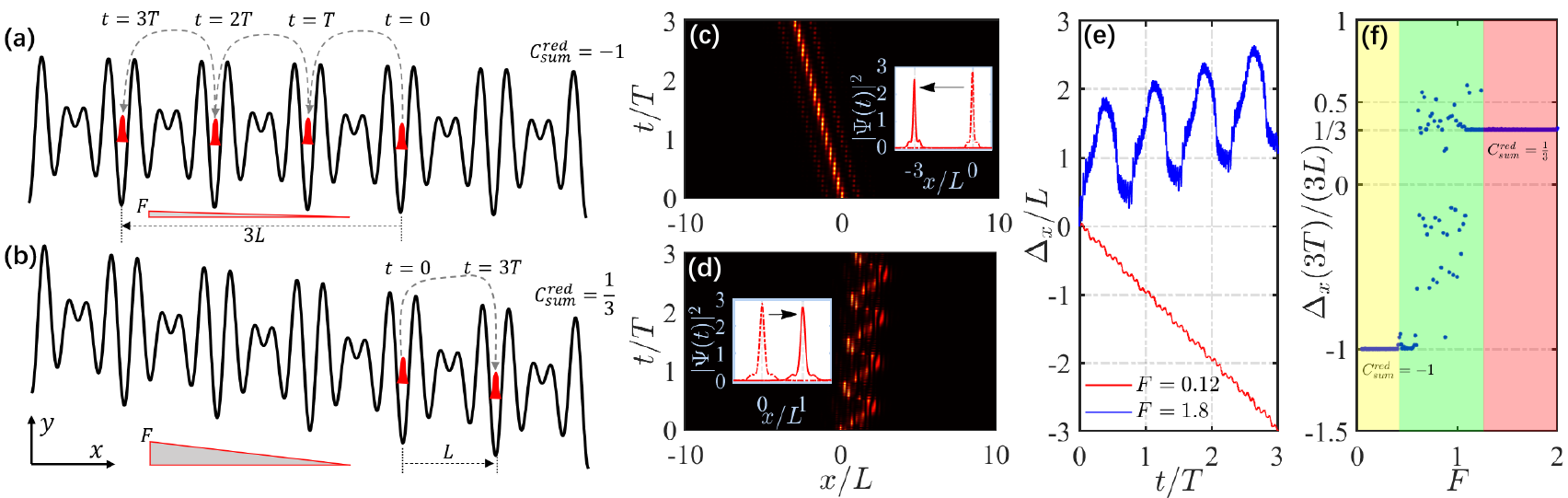}
\caption{Tilt-induced integral and fractional quantized-drift in a time-modulated superlattice with a gradient potential. (a) The weak tilt strength will cause integral quantized-drift to the left, where $F$ is the strength of tilt, $T$ is the period of time, $L$ is the period of superlattice, and $C_{sum}^{red}$ is the reduced expression of the Chern number. (b) The time-evolution of the wave-function profile in real space with $F=0.12$. The illustration describe the corresponding density distribution profile of initial
(dotted line) and final states (solid line), respectively. (c) The large tilt strength will cause fractional quantized-drift to the right. (d) The time-evolution of the wave-function profile in real space with $F=1.8$. (e) Drift $\Delta_x/L$ versus time $t$. (f) One-cycle-averaged transverse displacement $\Delta_x(3 T)/(3 L)$ as a function of the tilt strength $F$.
The other parameters are chosen as $\tau_1=\tau_2=25$, $d_1=1/2, d_2=2/3$ and $\omega=0.01$.
}\label{Fig2}
\end{figure*}

Adiabatic evolution makes it meaningful to consider the instantaneous spectrum of $\hat{H}(k,t)$, $\hat{H}(k,t) |\mu_{n}(k,t)\rangle= \varepsilon_{n} |\mu_{n}(k,t)\rangle$, where $|\mu_{n}(k,t)\rangle$ is the $n$th instantaneous eigenstate of $\hat{H}(k,t)$ and $\varepsilon_{n}$ is the corresponding instantaneous eigenenergy. This
gives origin to the instantaneous band gap spectrum, as shown in Figs.\ref{Fig1s}(a)and (b).
The complete two-dimensional energy band in first Brillouin zone is given in Fig.\ref{Fig1s}(a), where thin solid line represents the uniform sampling of the two-dimensional energy bands.
By expanding the sampling energy along the time direction, one can obtain equivalent one-dimensional time-dependent energy bands, see Fig.\ref{Fig1s}(b). The Chern numbers of corresponding first, second, third and fourth energy bands are $(C_1,C_2,C_3, C_4)=(-1,3,-1,-1)$.
It is found that we can use these one-dimensional time-dependent energy bands not only to accurately define the topological invariant Chern number, but also establish a close relationship between the drift displacement of wave packet and Chern number of the single or multiple bands.
The ratio of one-cycle drift displacement of wave packet to overall period of lattice is equivalent to the Chern number of single band, which will lead to integer drift.
The ratio of multi-cycle drift displacement of wave packet to multiple overall periods of lattice is equivalent to the equal probability superposition of Chern numbers of multiple bands, which will lead to the one-cycle-averaged drift displacement being fractional.

We consider the coordinate of the center of mass of the Gaussian wave packet defined as $X(t)=\int_{-\infty}^{+\infty} x |\Psi(x,t)|^2 d x$, and the initial Gaussian wave packet $\Psi(x,0)$ centered in $x_0$ with arbitrary mean quasimomentumat $k_0$ in the lowest energy band, $\Psi(x,0)=\zeta e^{-\frac{(x-x_0)^2}{4 D^2}}\widetilde{\mu}_{x}^n(k_0,0) e^{i k_0 x}$,
where $\zeta$ is a normalization factor, $D$ is the initial wave packet width, and $\widetilde{\mu}_{x}^n(k_0,0)$ is the amplitude of the real-space representation of the Bloch state $|\mu_n(k_0,0)\rangle$.
The initial Gaussian wave packet can be prepared by applying an additional harmonic trap~\cite{Lu2016Geometrical}.
In the following calculations, the time evolution of the wave packet is given by $\Psi(x,t)= \exp{(-iHt)}\Psi(x,0)$.
We then examine the density distribution profile $|\Psi(x,t)|^2$ of the time-evolving wave packet, and the mean drift displacement in $x$ direction given by $\Delta_x(t)=X(t)-X(0)$.
Interestingly, the mean drift displacement and direction of the wave packet after integer period of evolution are related to the strength of tilt.
A typical example is displayed
in Fig.\ref{Fig2}.
It can be clearly seen that
weak strength of tilt causes the drift of wave packet to the left, as shown in Figs.\ref{Fig2}(a) and (c). The large tilt strength causes the drift of wave packet to the right, see Figs.\ref{Fig2}(b) and (d).
One can obtain the mean drift displacement $\Delta_x{(3T)}/(3L)=-1$ for weak strength $F=0.12$ and $\Delta_x{(3T)}/(3L)=1/3$ for big strength $F=1.8$, see Fig.\ref{Fig2}(e). Obviously, the bigger tilt strength can induce fractional drift. It is worth noting that as the tilt strength increase, the integral or fractional quantization drifts occur in a platform way, see Fig.\ref{Fig2}(f).

In addition, because the tilt potential maybe breaks the adiabatic transport, the Landau Zener tunneling between two smaller band gaps is inevitable for a bigger tilt strength.
To do this, we characterize the Landau-Zener transition via using the occupation probability of $n$th bands, $P_{n}(t)=|\langle \psi(t)| \mu_{n}(k,t)\rangle|^2$.
According to the
theorem of adiabatic transport, the group velocity of
the particle along the $x$ direction for the momentum $k$ involved in the multi-bands comes from the energy dispersion and the Berry curvature
\begin{eqnarray} \label{Eq12}
v_g(k, t)=\sum_{n=1}^{\infty}P_n(t)[\frac{\partial \varepsilon_n(k, t)}{\partial k}+\mathcal{F}_n(k, t)],
\end{eqnarray}
where the Berry curvature is given by
\begin{eqnarray} \label{Eq7}
\mathcal{F}_n(k, t)=-2 \operatorname{Im}\left[\sum_{n^{\prime} \neq n} \frac{\langle u_n|\partial_k \hat{H}| u_{n^{\prime}}\rangle \langle u_{n^{\prime}} |\partial_t \hat{H} | u_n \rangle}{ (\varepsilon_n-\varepsilon_{n^{\prime}} )^2}\right]. \nonumber
\end{eqnarray}
Because the energy is a periodic function of time, the first term of group velocity $v_g(k, t)$ will periodically oscillate and hence induce
the conventional Bloch oscillation. Then the nontrivial Berry
curvature can induce quantized drift in the Bloch oscillation.
Then the Chern number of the $n$th band can be given by  $C_n=\frac{1}{2 \pi} \int_{-\pi / L}^{\pi / L} d k \int_0^{T}d t \mathcal{F}_n(k, t)$ (See section B of supplemental material).

If we consider a Bloch state of any quasi-momentum $k_0$ in the lowest energy band as the initial state,
the amount of drift displacement at time $\tau$ is simply given by the semiclassical expression $\Delta_x(\tau)=\int_0^\tau v_g(k_0, t) d t$.
This expression can be viewed as the time integral of the quantum flux determined by the group velocity $v_g$.
Because the occupation probability and instantaneous energy eigenvalues are periodic functions of time and momentum, the integral of the dispersion velocity in an overall period $T$ is exactly zero.
Thus the transverse drift displacement $\Delta_x$ over the duration of $\xi T$ contributed by  Berry curvatures of the multiple bands is
\begin{eqnarray} \label{Eq13}
\Delta_x(\xi T)&=&\sum_{n=1}^{\infty}\int_0^{\xi T} P_n(t) \mathcal{F}_n(k_0, t) d t \nonumber \\
&\simeq& \sum_{n=1}^{\infty}\bar{P}_{n}(\xi T)\int_0^{\xi T}\mathcal{F}_n(k_0, t) d t,
\end{eqnarray}
where $\xi$ denotes the number of energy bands involved in Landau Zener tunneling, and
$\bar{P}_{n}(t)=\frac{1}{\xi T}\int_0^{t}P_{n}(t)dt$ is the average occupation probability of the $n$-th band during the total time.
If the energy bands involved in Landau Zener tunneling exhibit an average population distribution, combining the condition $T_m/T_F\rightarrow0$ or $\gg1$ we can obtain the reduced expression of the Chern number (See section B of supplemental material)
\begin{eqnarray} \label{Eq14}
C_{sum}^{red}=\frac{\Delta_x(\xi T)}{qL\xi}\simeq \frac{1}{\xi}\sum_{n=1}^{\xi} C_n.
\end{eqnarray}
Obviously, $C_{sum}^{red}$ is effectively quantized integer or fraction because it is always very close to the ratio of the sum of Chern numbers of multiple bands to the number of energy bands involved in Landau Zener tunneling.
Especially, the Eq.\eqref{Eq14} shows that the transverse drift displacement $\Delta_x(\xi T)$ is not dependent on the initial momentum $k_0$. The main reason is that the integral value of Berry curvature $\int_{0}^{\xi T}\mathcal{F}_n(k_0, t)dt$ has strong robustness for selecting different quasi-momenta $k_0$ (See section C of supplemental material).

\begin{figure}[htp]
\center
\includegraphics[width=\columnwidth]{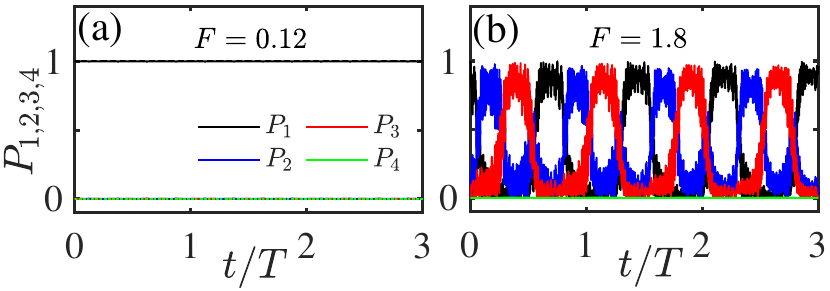}
\caption{(a) and (b) The corresponding occupation probabilities of the lowest four energy bands for $F=0.12$ and $1.8$, respectively. The other parameters are chosen as $\tau_1=\tau_2=25$, $d_1=1/2, d_2=2/3$, $w=0.01$ and $\xi=3$.
}\label{Fig3s}
\end{figure}

In Fig.\ref{Fig1s}(b), one can clearly see that the gaps between the lowest three energy bands are very close, which means that Landau-Zener tunneling is more likely to occur between these three energy bands.
To suppress Landau-Zener transition, the system needs to take a shorter time to travel across the anti-crossing point relative to
the tunneling time $T_{lz}=\sqrt{\vartheta} \max(1, \sqrt{\vartheta})/\Delta$, where $\vartheta=\Delta^2/(4\sqrt{\omega^2+F^2})$ is the adiabatic
parameter and $\Delta$ is the minimal energy gap~\cite{2010Landau,2020Nonlinear}.
Obviously, the weak strength of tilt can ensure well adiabatic evolution for fixing driving frequency $\omega$.
It is clear that the occupation probability $P_1$ always equal to one, which means
the system keeps staying in its instantaneous eigenstate of the first band
during the evolution for weal tilt strength, see Fig.\ref{Fig3s}(a).
Thus, combining the Eq.\eqref{Eq14} and $\xi=1$, one can obtain $C_{sum}^{red}=\Delta_x(T)/(L)= C_1=-1$, which is consistent with the result of dynamic evolution in Fig.\ref{Fig2}(e).
Similarly,
one can also obtain $C_{sum}^{red}=\Delta_x(T)/(L)=C_{2,3}=3$ or $-1$ for initially sweeping from  the second or third energy band, respectively.
On the contrary, for a large tilt strength, we can observe Rabi oscillations
between three lowest bands due to occurrence of Landau-Zener tunneling, see Fig.\ref{Fig3s}(b).
Interestingly, It is found that the average probability of occupying three lowest bands is equal after integer period $3 T$, that is $\bar{P}_n(3 T)=\frac{1}{3 T}\int_0^{3T}P_{n}(t)dt=1/3$,
for $n=1,2,3$.
Thus, according to the Eq.\eqref{Eq14} and $\xi=3$, one can obtain Chern number $C_{sum}^{red}=\Delta_x(3T)/(3L)=(C_1+C_2+C_3)/3=1/3$, which means that the  one-cycle-averaged drift displacement is fraction $\frac{1}{3}$, consisting with the result of dynamic evolution in Fig.\ref{Fig2}(e). Meanwhile, it also provide one method of measuring topology invariant of energy band via quantized drift.

\begin{figure}[htp]
\center
\includegraphics[width=\columnwidth]{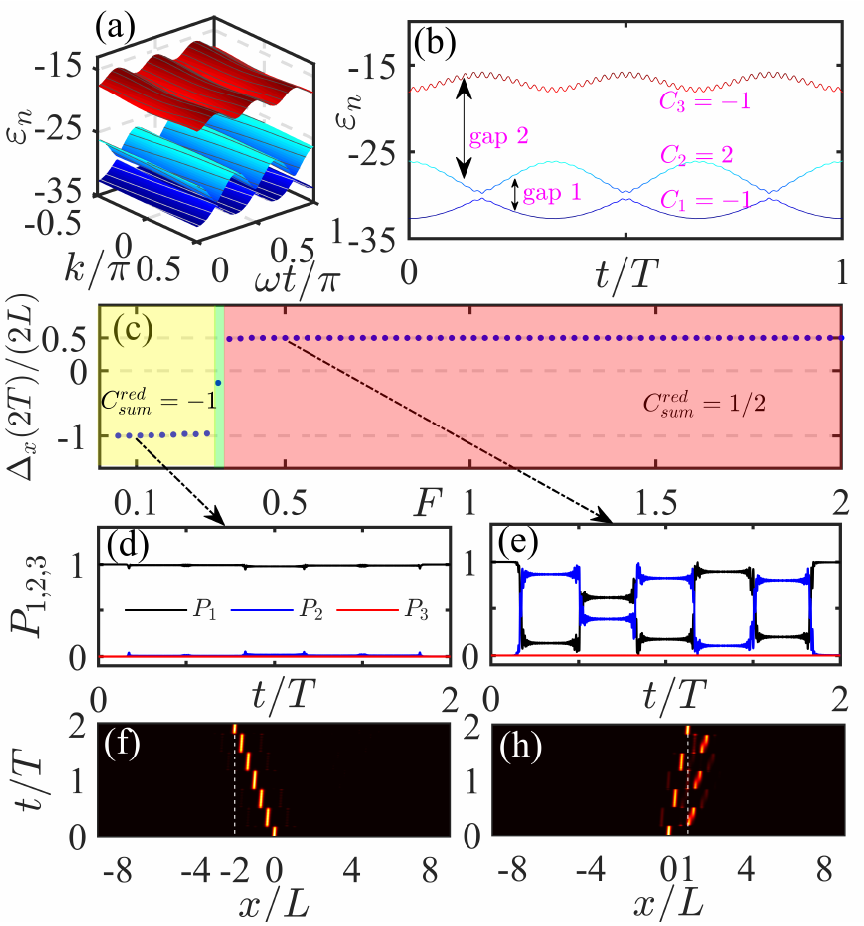}
\caption{(a) The complete two-dimensional energy band in the first Brillouin zone, where thin solid line represents the uniform sampling of the two-dimensional energy bands in the presence of tilt. (b) The equivalent one-dimensional time-dependent energy bands. (c) One-cycle-averaged drift displacement $\Delta_x(2 T)/(2 L)$ as a function of the tilt strength $F$. (d) and (e) The corresponding probability population of the lowest three
energy bands with $F = 0.1$ and $0.5$, respectively. (f) and (h) The corresponding time-evolution of the wave-function in real space. The other parameters are chosen as $\tau_1=\tau_2=25$, $d_1=2/3, d_2=1$, $w=0.005$ and $\xi=2$.
}\label{Fig3}
\end{figure}

Our system also can be used to produce fractional drift with other fraction by adjusting period of
lattice $d_1$ and $d_2$, such as $1/2$, $1/4$ and so on.
As an example, we consider the case of $d_1=2/3$ and $d_2=1$ in Fig.\ref{Fig3}, which can produce fractional drift with $\frac{1}{2}$, the other parameters are consistent with the previous discussion.
An inspection of the two-dimensional energy band and one-dimensional time-dependent energy band  diagram shows that, the gap between the lowest two bands is very close, which means that  Landau-Zener tunneling is more likely to occur between the lowest two bands, see Figs.\ref{Fig3}  (a) and (b). The weak strength of tilt
can ensure well adiabatic evolution, and lead to integral drift $C_{sum}^{red}=\Delta_x(T)/(L)= C_1=-1$, as shown in Figs.\ref{Fig3}(c), (d) and (f).
Obviously, for a large tilt strength, we can observe Rabi
oscillations between two lowest bands due to occurrence
of Landau-Zener tunneling, and the average probability of occupying two
lowest bands is equal after integer period $2 T$.
Thus, one can obtain Chern number $C_{sum}^{red}=\Delta_x(2T)/(2L)=(C_1+C_2)/2=1/2$, which means that the one-cycle-averaged drift displacement is fraction $\frac{1}{2}$ [see Figs.\ref{Fig3}(c), (e) and (h)].
%

In summary, we have studied tilt-induced fractional quantize drift in non-interacting system.
If the tilt strength is weak, the system can ensure well adiabatic evolution, and keep
staying in its instantaneous eigenstate of the initial energy band, one-cycle-averaged drift displacement is integer that is relate to Chern number of this energy band.
For bigger tilt strength, Rabi oscillations between adjacent lowest enegy bands caused by
Landau-Zener tunneling, induce that the one-cycle-averaged drift displacement is fraction,
which is relate to the ratio of the sum of Chern numbers of multiple bands to the
number of energy bands involved in Landau Zener tunneling.

In addition to the discovery of the tilt-induced fractional quantize drift,
we believe that our work brings three key advances to
related fields. First, different from fractional quantize response induced by nonlinearity that
is difficult to manipulate in experiment, the tilt strength controlled by gradient magnetic field
more easily adjust in experiment, which may stimulate experimental realization of fractional quantize response in various systems, including optical lattice, optical waveguide, acoustic system. Second, tilt-driven the system to turn from integral quantize drift to fractional quantize drift provide a new route for measuring topology invariant of energy band. Third, as different  fractional drift with different fraction can be effectively designed by adjusting period of lattice, and the propagating direction and displacement of the corresponding intensity can be controlled, this offers an efficient way to beam controller
in optical systems.
%

\begin{acknowledgments}
The authors are thankful to C. Lee and Y. Ke for enlightening suggestions
and helpful discussions.
This work is supported by the National Natural Science
Foundation of China under Grant No.12205385 and No.12175315, the Scientific Research Fund of Hunan Provincial Education Department under Grant No. 23B0264, and the Talent Project of Central South University of Forestry and Technology under Grant No.2021YJ025.
\end{acknowledgments}

\bibliography{quantum_quench}

\appendix

\section{The Hamiltonian in momentum space } \label{Appendix1}
For the system without tilt ($V_1(x)=0$), the Schr\"{o}dinger equation satisfied by the instantaneous Hamiltonian can be written as
\begin{eqnarray} \label{Ham1z}
i\frac{\partial \psi(x,t)}{\partial t}=-\frac{\partial^2}{\partial x^2} \psi(x,t) +V_0(x,t).
\end{eqnarray}
Because the superlattice potential $V_0$ is a periodic function with respect to coordinate $x$, which the overall period is $L=n_1d_2=n_2d_1$,
the wave function $\psi(x,t)$ can be expanded into a linear superposition of the Bloch function $\varphi_{k,n}(x)$,
\begin{eqnarray} \label{Ham2z}
\psi(x,t)=\sum_{k,n}a_{k,n}\varphi_{k,n}(x)e^{-i\varepsilon_{k,n}t},
\end{eqnarray}
where $k$ is the quasi-momentum, and $\varepsilon_{k,n}$ is the instantaneous energy of $n$th band.
Substituting equation~\eqref{Ham2z} into equation~\eqref{Ham1z}, one can obtain the eigenvalue equation
\begin{eqnarray} \label{Ham3z}
\left[-\frac{1}{2}\frac{\partial^2}{\partial x^2}+V_0(x,t)\right]\varphi_{k,n}(x)=\varepsilon_{k,n}\varphi_{k,n}(x).
\end{eqnarray}
We assume that the Bloch function for $n$th band is a superposition of plane waves with different quasi-momenta $k$,
\begin{eqnarray} \label{Ham4z}
&&\varphi_{\mathrm{k}, n}(x)=\sum_{j=-N}^N c_{\mathrm{k}, n, j} e^{i\left(k+\frac{2 \pi}{L} j\right) x}, \nonumber \\
&&\varphi_{\mathrm{k}, n}(x+L)=\sum_{\mathrm{j}=-N}^N c_{k, n, j} e^{i(k+\frac{2 \pi}{L} j)(x+L)},
\end{eqnarray}
where $k\in[-\pi/L,\pi/L]$.
Substituting the above Bloch function into the eigenvalue equation, we have
\begin{eqnarray} \label{Ham5z}
&&\frac{1}{2}\sum_{j=-N}^{N}\left[\left(k+\frac{2 \pi}{L} j\right)^2-\left(\tau_1+\tau_2\right)\right] c_{k, n, j}\nonumber \\
&&-\frac{\tau_2}{4}\sum_{j=-N}^{N}\left(e^{i 2 \omega t} c_{k, n, j-\frac{L}{d_2}}+e^{-i 2 \omega t} c_{k, n, j+\frac{L}{d_2}}\right) \nonumber \\
&&-\frac{\tau_1}{4}\sum_{j=-N}^{N}\left(c_{k, n, j-\frac{L}{d_1}}+c_{k, n, j+\frac{L}{d_1}}\right)=\varepsilon_{\mathrm{k}, n}\sum_{j=-N}^{N} c_{k, n, j}. \nonumber
\end{eqnarray}
Using expansion coefficient $\{c_{k, n, j}\}$ with $j=-N,\cdots,0,\cdots,N$ as the base vector, the above eigenvalue equation can be written in matrix form, whose left item corresponds to the representation of the instantaneous Hamiltonian in momentum space $H (k,t)$.
By rewriting $c_{k, n, j}$ as operator $\hat{c}_j$,
the instantaneous Hamiltonian in momentum space $H (k,t)$ can be written in the form of second quantization,
\begin{eqnarray} \label{Ham6z}
\hat{H}(k,t)&=&-\sum_j\left[(\frac{\tau_2}{4}e^{i 2 \omega t}) \hat{c}_j^{\dagger} \hat{c}_{j-\frac{L}{d_2}}+(\frac{\tau_1}{4}) \hat{c}_j^{\dagger} \hat{c}_{j-\frac{L}{d_1}}+\text { H.c. }\right] \nonumber \\
&+&\frac{1}{2}\sum_j\left[\left(k+\frac{2 \pi}{L} j\right)^2-\left(\tau_1+\tau_2\right)\right] \hat{n}_j,
\end{eqnarray}
where $\hat{c}^{(\dagger)}_j$ is the annihilation (creation) operator of lattice $j$ in the $n$th band, and $\hat{n}_j=\hat{c}_j^{\dagger} \hat{c}_{j}$ is the density operator.

In the presence of the tilt ($V_1(x)=Fx$), we can equivalently deal with the problem in a rotational framework by making a unitary transformation $\Psi(x,t)=e^{-iFxt}\psi(x,t)$.
The instantaneous Hamiltonian in the form of second quantization can be obtained by replacing $k$ with $k-Ft$,
\begin{small}
\begin{eqnarray} \label{Ham7z}
\hat{H}(k,t)&=&-\sum_j\left[(\frac{\tau_2}{4}e^{i 2 \omega t}) \hat{c}_j^{\dagger} \hat{c}_{j-\frac{L}{d_2}}+(\frac{\tau_1}{4}) \hat{c}_j^{\dagger} \hat{c}_{j-\frac{L}{d_1}}+\text { H.c. }\right] \nonumber \\
&+&\frac{1}{2}\sum_j\left[\left(k-Ft+\frac{2 \pi}{L} j\right)^2-\left(\tau_1+\tau_2\right)\right] \hat{n}_j.
\end{eqnarray}
\end{small}
If we replace ($k-Ft$) and $\omega t$ with $K_x$ and $K_y$ respectively, the instantaneous Hamiltonian~\eqref{Ham7z} can be regarded as an equivalent static Hamiltonian
\begin{small}
\begin{eqnarray} \label{s_Ham5}
\hat{H}(K_x,K_y)&=&-\sum_j\left[(\frac{\tau_2}{4}e^{i 2 K_y}) \hat{c}_j^{\dagger} \hat{c}_{j-\frac{L}{d_2}}+(\frac{\tau_1}{4}) \hat{c}_j^{\dagger} \hat{c}_{j-\frac{L}{d_1}}+\text { H.c. }\right] \nonumber \\
&+&\frac{1}{2}\sum_j\left[\left(K_x+\frac{2 \pi}{L} j\right)^2-\left(\tau_1+\tau_2\right)\right] \hat{n}_j, \nonumber \\
\end{eqnarray}
\end{small}
where $K_x\in[-\pi/L,\pi/L]$ and $K_y\in[0,\pi]$ can be understood as the quasi-momenta of the two-dimensional first Brillouin zone.
%

\section{The relation between $C_{sum}^{red}$ and $C_n$} \label{Appendix2}
According to the equivalent static Hamiltonian~\eqref{s_Ham5}, the conventional Chern number for the $n$th energy band can be defined by the integral of Berry curvature in the two-dimensional  Brillouin zone as
\begin{eqnarray} \label{s_Ham6}
C_n=\frac{1}{2\pi}\int_{-\pi/L}^{\pi/L}dK_x\int_{0}^{\pi}dK_y \mathcal{F}_n(K_x, K_y),
\end{eqnarray}
where the Berry curvature is given by
\begin{widetext}
\begin{eqnarray}\label{s_Ham7}
\mathcal{F}_n(K_x, K_y)&=&-2 \operatorname{Im}\left[\sum_{n^{\prime} \neq n} \frac{\langle u_n|\partial_{K_x} \hat{H}| u_{n^{\prime}}\rangle \langle u_{n^{\prime}} |\partial_{K_y} \hat{H} | u_n \rangle}{ (\varepsilon_n-\varepsilon_{n^{\prime}} )^2}\right]
=-2 \operatorname{Im}\left[\sum_{n^{\prime} \neq n} \frac{\langle u_n|\partial_{K_x} \hat{H}| u_{n^{\prime}}\rangle \langle u_{n^{\prime}} |(\frac{\partial_{t}}{\partial_{K_y}})\partial_{t} \hat{H} | u_n \rangle}{ (\varepsilon_n-\varepsilon_{n^{\prime}} )^2}\right] \nonumber \\
&=&-\frac{2}{\omega} \operatorname{Im}\left[\sum_{n^{\prime} \neq n} \frac{\langle u_n|\partial_{K_x} \hat{H}| u_{n^{\prime}}\rangle \langle u_{n^{\prime}}| \partial_{t} \hat{H} | u_n \rangle}{ (\varepsilon_n-\varepsilon_{n^{\prime}} )^2}\right]
=\frac{1}{\omega}\mathcal{F}_n(K_x,t)=\frac{1}{\omega}\mathcal{F}_n(k, t).
\end{eqnarray}
\end{widetext}
Then the Chern number can be further rewritten as
\begin{eqnarray} \label{s_Ham8}
C_n&=&\frac{1}{2\pi}\int_{-\pi/L}^{\pi/L}dK_x\int_{0}^{\pi}dK_y \mathcal{F}_n(K_x, K_y) \nonumber \\
&=&\frac{1}{2\pi}\int_{-\pi/L}^{\pi/L}dK_x\int_{0}^{T_m}dt \mathcal{F}_n(K_x, t) \nonumber \\
&=&\frac{1}{2\pi q}\int_{-\pi/L}^{\pi/L}dK_x\int_{0}^{T}dt \mathcal{F}_n(K_x, t)\nonumber \\
&=&\frac{1}{2\pi q}\int_{-\pi/L}^{\pi/L}dk\int_{0}^{T}dt \mathcal{F}_n(k, t).
\end{eqnarray}
%
When $T_m/T_F\rightarrow0$ or $\gg1$, the integral $\int_{0}^{T}dt \mathcal{F}_n(k, t)$ is independent of the quasi-momentum $k$ (see Appendix C for detail).
Hence we can get rid of the integral on $k$ in the above equqtion and obtain
\begin{eqnarray} \label{s_Ham9}
C_n&=&\frac{1}{2\pi q}\int_{-\pi/L}^{\pi/L}dk\int_{0}^{T}dt \mathcal{F}_n(k, t) \nonumber \\
&=&\frac{1}{ qL}\int_{0}^{T}dt \mathcal{F}_n(k_0, t) \nonumber \\
&=&\frac{1}{qLm}\int_{0}^{m T}dt \mathcal{F}_n(k_0, t),
\end{eqnarray}
where $k_0$ and $m$ represent an arbitrary mean quasi-momentum and positive integer, respectively.

For a Bloch state with quasi-momentum $k_0$ involved in multi-bands, the mean
displacement in $x$ direction $\Delta_x(t)=X(t)-X(0)$ at time $t$ can be
given by the semiclassical expression
\begin{eqnarray} \label{s_Ham10}
\Delta_x(\tau)=\int_0^\tau v_g(k_0, t) d t,
\end{eqnarray}
with
\begin{eqnarray}
v_g(k_0, t)=\sum_{n=1}^{\infty}P_n(t)[\frac{\partial \varepsilon_n(k_0, t)}{\partial k}+\mathcal{F}_n(k_0, t)],
\end{eqnarray}
where $P_n(t)$ is the occupation probability of
$n$th band.
Because the occupation probability and instantaneous energy are periodic functions of time, the first term of group velocity $v_g(k_0, t)$ will periodically oscillate with time, and then the integral of dispersion velocity
is exactly zero in the overall period $T$.
In addition, the common period doubling of the occupation probability $P_n(t)$ and Berry curvature $\mathcal{F}_n(k_0, t)$ is $\xi T$.
Thus the drift displacement $\Delta_x$ over the duration of $\xi T$ contributed by the multi-bands Berry curvatures is
\begin{eqnarray} \label{s_Ham11}
\Delta_x(\xi T)&=&\sum_{n=1}^{\infty}\int_0^{\xi T} P_n(t) \mathcal{F}_n(k_0, t) d t \nonumber \\
&=& \sum_{n=1}^{\infty}\frac{1}{\xi T}\int_0^{\xi T} P_n(t)d t\int_0^{\xi T}\mathcal{F}_n(k_0, t) d t \nonumber \\
&=& \sum_{n=1}^{\infty}\bar{P}_n(\xi T)\int_0^{\xi T}  \mathcal{F}_n(k_0, t) d t,
\end{eqnarray}
where $\xi$ denotes the number of energy bands involved in Landau Zener tunneling, and
$\bar{P}_{n}(t)=\frac{1}{\xi T}\int_0^{t}P_{n}(t)dt$ is the average occupation probability of the $n$-th band during the total time.
If the energy bands involved in Landau Zener tunneling exhibit an average population uniform distribution, $\bar{P}_{n}(\xi T)=1/\xi$, we can obtain
\begin{eqnarray} \label{s_Ham12}
\Delta_x(\xi T)
&\simeq & \sum_{n=1}^{\infty}\bar{P}_n(\xi T) \int_0^{\xi T}  \mathcal{F}_n(k_0, t) d t \nonumber \\
&=& \frac{1}{\xi}\sum_{n=1}^{\xi}\int_0^{\xi T} \mathcal{F}_n(k_0, t) d t.
\end{eqnarray}
Combining the equation~\eqref{s_Ham9}, the equation~\eqref{s_Ham12} can be rewritten as
\begin{eqnarray} \label{s_Ham13}
C_{sum}^{red}=\frac{\Delta_x(\xi T)}{q L \xi}
&= & \frac{1}{\xi} \frac{1}{q L \xi} \sum_{n=1}^{\xi}\int_0^{\xi T} \mathcal{F}_n(k_0, t) d t \nonumber \\
&=& \frac{1}{\xi}\sum_{n=1}^{\xi}C_n.
\end{eqnarray}
$C_{sum}^{red}$ can be effectively regarded as a reduced expression of Chern number, in which  multi-bands involved in Landau Zener tunneling are equal probability occupation.

\begin{figure}[htp]
\center
\includegraphics[width=\columnwidth]{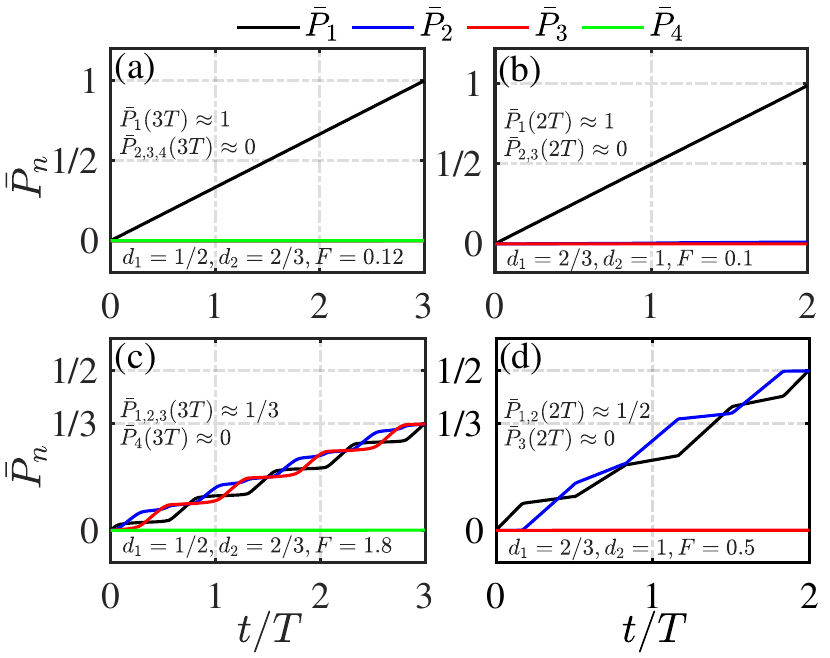}
\caption{The average occupation probability $\bar{P}_n(t)$ as a function of time for different tilt strengths. (a) and (c) The superlattice potential has fractional
drift $1/3$. (b) and (d) The superlattice potential has fractional
drift $1/2$. The other parameters are chosen as $\tau_1=\tau_2=25$.
}\label{Fig5}
\end{figure}

To numerically verify the equal probability occupation for multi-bands involved in Landau Zener tunneling, we give the average probability $\bar{P}_n(t)$ as a function of time for different tilt strengths by preparing the initial state on the lower-energy band, as shown in Fig.\ref{Fig5}.
For the cases of weak tilt strengths, the system can adiabatically follow the instantaneous eigenstate of the first band, and the average probability $\bar{P}_n(\xi T)$ always equal to $1$, see Figs.\ref{Fig5}(a) and (b).
For the case of large tilt strengths, Rabi oscillations can occur in the lowest adjacent energy bands due to Landau-Zener tunneling, and result in an equal probability occupation for multi-bands involved in Landau Zener tunneling, $\bar{P}_n(3 T)=1/3$ (Fig.\ref{Fig5}(c)) and $\bar{P}_n(2  T)=1/2$ (Fig.\ref{Fig5}(d)).

\section{The dependence of the integral of the Berry curvature on quasi-momentum $k$ }   \label{Appendix3}

In this section, we show the one-dimensional time integral of the Berry curvature, $\int_{0}^{T}dt \mathcal{F}_n(k, t)$, is independent of the initial momentum value of a Bloch
state $k$ when $T_m/T_F \rightarrow 0$ or $\gg1$.
Because the Berry curvature $\mathcal{F}_n(k, t)$ is a periodic function of quasi-momentum and time,
the integral of Berry curvature over an overall time period is independent of the initial time.
In addition, the quasi-momentum $k$ change with time as $k-Ft$, and then it is equivalent to shift $k$ to $k+\Delta k$ while it maintains the time as $t$.
We can obtain
\begin{eqnarray} \label{s_Ham14}
\int_{0}^{T} \mathcal{F}_n(k, t)dt&=&\omega \int_{0}^{\frac{T}{\omega}} \mathcal{F}_n(k, \omega t)dt \nonumber \\
&=&\omega \int_{\frac{\Delta k}{F}}^{\frac{T}{\omega}+\frac{\Delta k}{F}} \mathcal{F}_n(k+\Delta k,   \omega (t-\frac{\Delta k}{F}))dt. \nonumber \\
\end{eqnarray}
If $T_m/T_F\gg 1$, we have $\frac{\Delta k \omega}{F}=\frac{\Delta k L T_F}{2 T_m}\simeq 0$, and obtain
\begin{eqnarray} \label{s_Ham16}
\int_{0}^{T} \mathcal{F}_n(k, t)dt&=&\omega \int_{0}^{\frac{T}{\omega}} \mathcal{F}_n(k, \omega t)dt \nonumber \\
&=&\omega \int_{\frac{\Delta k}{F}}^{\frac{T}{\omega}+\frac{\Delta k}{F}} \mathcal{F}_n(k+\Delta k, \omega (t-\frac{\Delta k}{F}))dt \nonumber \\
&=&\omega \int_{\frac{\Delta k}{F}}^{\frac{T}{\omega}+\frac{\Delta k}{F}} \mathcal{F}_n(k+\Delta k, \omega t)dt \nonumber \\
&=&\int_{\frac{\Delta k \omega}{F}}^{T+\frac{\Delta k \omega}{F}} \mathcal{F}_n(k+\Delta k, t)dt. \nonumber \\
&=&\int_{0}^{T} \mathcal{F}_n(k+\Delta k, t)dt.
\end{eqnarray}
If $T_m/T_F\rightarrow 0$, we have $\frac{\Delta k' F}{\omega}=\frac{2 \Delta k' T_m}{L T_F}\simeq 0$, and obtain
\begin{eqnarray} \label{s_Ham16}
\int_{0}^{T} \mathcal{F}_n(k, t)dt&=&\omega \int_{0}^{\frac{T}{\omega}} \mathcal{F}_n(k, \omega t)dt \nonumber \\
&=&\omega \int_{\frac{\Delta k'}{\omega}}^{\frac{T}{\omega}+\frac{\Delta k'}{\omega}} \mathcal{F}_n(k+\frac{\Delta k' F}{\omega}, \omega (t-\frac{\Delta k'}{\omega}))dt \nonumber \\
&=&\omega \int_{\frac{\Delta k'}{\omega}}^{\frac{T}{\omega}+\frac{\Delta k'}{\omega}} \mathcal{F}_n(k, \omega (t-\frac{\Delta k'}{\omega}))dt \nonumber \\
&=&\int_{\Delta k'}^{T+\Delta k'} \mathcal{F}_n(k, t-\Delta k')dt. \nonumber \\
&=&\int_{0}^{T} \mathcal{F}_n(k-F\Delta k', t)dt.
\end{eqnarray}
Thus the one-dimensional time integral of the Berry curvature is independent of quasi-momentum $k$, if $T_m/T_F \rightarrow 0$ or $\gg1$.

\end{document}